\def\be{\begin{equation}}
\def\ee{\end{equation}}
\def\bea{\begin{eqnarray}}
\def\eea{\end{eqnarray}}
\def\bse{\begin{subequations}}
\def\ese{\end{subequations}}

\documentclass[prb,twocolumn,showpacs,amsmath,amssymb,eqsecnum,preprintnumbers]{revtex4}
\usepackage{graphicx}
\usepackage{dcolumn}
\usepackage{bm}
\begin{document}
\title{Phase-ordering dynamics in itinerant quantum ferromagnets}


\author{D. Belitz}
\affiliation{Department of Physics and Institute for Theoretical Science,
              University of Oregon, Eugene, OR 97403
          }
\author{T.R. Kirkpatrick}
\affiliation{Institute for Physical Science and Technology and
                    Department of Physics, University of Maryland,
                    College Park, MD 20742
          }
\author{Ronojoy Saha}
\affiliation{Institute for Physical Science and Technology and Department of
             Physics, University of Maryland, College Park, MD 20742\\
             and\\
             Department of Physics and Materials Science Institute,
             University of Oregon, Eugene, OR 97403
          }
\date{\today}
\begin{abstract}
The phase-ordering dynamics that result from domain coarsening are considered
for itinerant quantum ferromagnets. The fluctuation effects that invalidate the
Hertz theory of the quantum phase transition also affect the phase ordering.
For a quench into the ordered phase there appears a transient regime where the
domain growth follows a different power law than in the classical case, and for
asymptotically long times the pre\-factor of the $t^{1/2}$ growth law has an
anomalous magnetization dependence. A quench to the quantum critical point
results in a growth law that is not a power-law function of time. Both
phenomenological scaling arguments and renormalization-group arguments are
given to derive these results, and estimates of experimentally relevant length
and time scales are presented.
\end{abstract}


\maketitle

\section{Introduction}
\label{sec:I}

When a many-body system capable of a phase transition from a disordered phase
to a phase with long-range order is suddenly taken, by changing one or more
parameters, from the disordered phase to the ordered one, an interesting
question is how the long-range order will develop as time goes by and the
system approaches equilibrium. Such a sudden transformation is called a
``quench'', and the phase-ordering dynamics after the quench can be studied by
kinetic methods similar to those used for the critical dynamics near the phase
transition.\cite{Hohenberg_Halperin_1977} The quenching problem is of broad
interest since it is applicable to a large variety of physical systems that
undergo phase transitions, ranging from magnets to liquid helium to the early
universe.\cite{Zurek_1996} The phase ordering occurs by means of the growth of
domains that arise from spontaneous fluctuations, and the linear size $L$ of
these domains\cite{domains_footnote} obeys a power-law as a function of time
$t$ for sufficiently large $t$: $L(t) \propto t^{1/z}$, with $z$ a dynamical
exponent.\cite{Bray_1994} In addition, the pair correlation function is
observed, both experimentally and numerically, to obey a simple scaling law
\be
C(r,t) \equiv \langle {\bm\phi}({\bm x},t)\cdot{\bm\phi}({\bm 0},t)\rangle
= f(r/L(t)),
\label{eq:1.1}
\ee
Here $r = \vert{\bm x}\vert$, $f$ is a scaling function, ${\bm\phi}$ is the
order parameter field, and $\langle \ldots \rangle$ denotes a statistical
average. We take the order parameter to be a real 3-vector, that is, we
consider Heisenberg magnets.

The facts stated above, although well established,\cite{Bray_1994} are purely
phenomenological; so far no derivation from first principles has been given.
This phenomenology has so far been applied to classical systems, but there is
no reason to expect that it will not be valid for quantum systems as well.
Quantum phase transitions are known to differ in crucial aspects from classical
ones,\cite{Sachdev_1999, Belitz_Kirkpatrick_Vojta_2005} and one needs to ask
whether these differences affect the phase ordering properties as well. In this
paper we investigate this problem for the case of a quantum ferromagnet and
show that the phase ordering kinetics are indeed affected in dramatic ways.

\section{Review of results for classical magnets}
\label{sec:II}

In order to motivate our approach and put it into context, we first briefly
recall the known results for phase ordering in classical magnets. The dynamical
equation that governs the time evolution of the order parameter in an isotropic
Heisenberg ferromagnet is\cite{Ma_Mazenko_1975}
\be
\frac{\partial{\bm\phi}}{\partial t} = \lambda\,{\bm\nabla}^2\,\frac{\delta
H}{\delta{\bm\phi}} + \gamma\,{\bm\phi}\times\frac{\delta H}{\delta{\bm\phi}} +
{\bm\zeta}. \label{eq:2.1}
\ee
Here $\lambda$ is a spin transport coefficient, and $\gamma$ is a gyromagnetic
ratio. The Langevin force ${\bm\zeta}$ is of no consequence for the problem of
phase ordering and we will neglect it.\cite{Bray_1994} $H$ is the Hamiltonian
or free energy functional that governs the equilibrium properties of the
system. A classical isotropic Heisenberg ferromagnet is described by a
$\phi^4$-theory with parameters $r$, $c>0$, and $u>0$\cite{Ma_1976}
\bse
\label{eqs:2.2}
\be H = \int d{\bm
x}\,\left[\frac{c}{2}\left(\nabla{\bm\phi}\right)^2 + \frac{r}{2}\,{\bm\phi}^2
+ \frac{u}{4}\,({\bm\phi}^2)^2\right].
\label{eq:2.2a}
\ee
In Fourier space, this corresponds to a Gaussian
vertex
\be \Gamma({\bm k}) = r + c\,{\bm k}^2.
\label{eq:2.2b}
\ee
\ese
The first term in Eq.\ (\ref{eq:2.1}) describes dissipative dynamics for a
conserved order parameter; this is Model B in Ref.\
\onlinecite{Hohenberg_Halperin_1977}. Including the second term takes into
account the precession of spins in the effective magnetic field created by all
other spins; this is Model J in Ref.\ \onlinecite{Hohenberg_Halperin_1977}.

The phase-ordering problem for Model B ($\gamma=0$) has been studied by a
variety of analytic techniques as well as by simulations.\cite{Bray_1994} The
result is a dynamical exponent $z=4$; i.e., the linear domain size grows for
long times as $L(t) \propto t^{1/4}$, with a pre\-factor that is independent of
the equilibrium magnetization $m_0$. This result is plausible from a simple
power-counting argument: Eq.\ (\ref{eq:2.1}) has the structure of a continuity
equation for each component $\phi_{\alpha}$ of the order parameter,
$\partial_t\,\phi_{\alpha} = -{\bm\nabla}\cdot{\bm j}_{\alpha}$, and for power
counting purposes the current ${\bm j}_{\alpha}$ can be identified with the
domain growth velocity $dL/dt$ times $m_0$. Assuming that each gradient can be
identified with a factor of $1/L$,\cite{Ising_footnote} this leads to $dL/dt
\sim c\lambda/L^3$,\cite{notation_footnote, gradients_footnote} or
\be
L(t) \propto (c\lambda)^{1/4}\,t^{1/4}\quad,\quad\text{(Model B)}.
\label{eq:2.3}
\ee

The additional term in Model J ($\gamma \neq 0$) describes spin waves whose
dispersion relation follows from Eq.\ (\ref{eq:2.1}):
\be
\omega({\bm k}) = D(m_0)\,{\bm k}^2,
\label{eq:2.4}
\ee
with $D(m_0) = \gamma\,c\,m_0 \equiv c\Omega_{\text{L}}$, where
$\Omega_{\text{L}} = \gamma\,m_0$ is the Larmor frequency related to the
equilibrium magnetization $m_0$. This is consistent with results obtained from
microscopic models.\cite{Moriya_1985} For the phase ordering problem, Model J
was studied in Ref.\ \onlinecite{Das_Rao_2000}. The same power-counting
arguments as for Model B above suggest
\be \frac{dL}{dt} \sim \frac{c\lambda}{L^3} + \frac{c\,\Omega_{\text{L}}}{L}\ .
\label{eq:2.5}
\ee
According to Eq.\ (\ref{eq:2.5}), the time dependence of $L$ will cross over
from the $t^{1/4}$ behavior characteristic for Model B to a $t^{1/2}$ behavior
at a length scale $L_1 = \sqrt{\lambda/\Omega_{\text{L}}}$,
\be
L(t) \propto \begin{cases} (c\lambda)^{1/4}\,t^{1/4} &
\text{if}\ L\ll L_1 \cr
                           (c\Omega_{\text{L}})^{1/2}\,t^{1/2} & \text{if}\ L\gg
                           L_1
              \end{cases} \quad,\quad\text{(Model J)}.
\label{eq:2.6}
\ee
A numerical solution of the dynamical equation was found to be in good
agreement with this expectation.\cite{Das_Rao_2000}

This concludes our review of known results for classical magnets.

\section{Quantum Ferromagnets}
\label{sec:III}

\subsection{Mode-mode coupling effects}
\label{subsec:III.A}

\begin{figure}[t,b]
\vskip -0mm
\includegraphics[width=8cm]{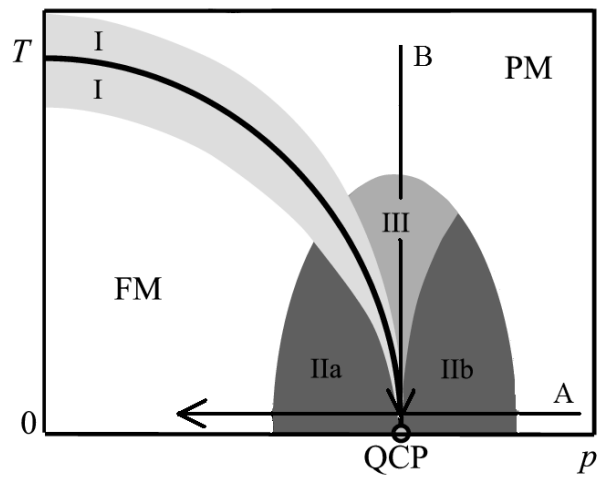}
\caption{Schematic phase diagram in a temperature (T) -
pressure (p) plane with a phase separation line separating a
paramagnetic (PM) phase from a ferromagnetic (FM) one. Shown are the
quantum critical point (QCP), the classical critical regime (I), and
static (IIa,b) and dynamic (III) quantum critical regimes. The
arrows denote a quench into the ordered phase (A), and a critical
quench (B), respectively. See the text for further explanation, and
Ref.\ \onlinecite{Belitz_Kirkpatrick_Vojta_2005} for a general
discussion of magnetic quantum criticality.}
\label{fig:1}
\end{figure}

The above results hold if the equilibrium properties of the ferromagnet are
described by Eqs.\ (\ref{eqs:2.2}), and if there are no other soft modes that
couple to the order parameter. A counterexample is phase separation in binary
fluids, where one needs to take into account that the local fluid velocity
contributes to the order parameter transport.\cite{Bray_1994} The net result is
equivalent to a nonlocal free energy, or dynamic equation, and this is obtained
explicitly if the additional soft modes are integrated out. At low temperature
($T$) in itinerant ferromagnets a similar phenomenon occurs; viz., a coupling
of the order parameter to soft particle-hole
excitations.\cite{Kirkpatrick_Belitz_1996, Belitz_et_al_2001a} These mode-mode
coupling effects invalidate Hertz's mean-field theory,\cite{Hertz_1976} and
they change both the critical behavior at the ferromagnetic quantum phase
transition,\cite{Belitz_et_al_2001b} and the magnetization dependence of the
magnon dispersion relation in the ordered phase.\cite{Belitz_et_al_1998} Here
we investigate how these effects influence the phase ordering following a
quench into the quantum regime, which can be realized by a pressure quench at
fixed low temperature in a system where the ferromagnetic quantum phase
transition can be tuned by hydrostatic pressure, see Fig.\ \ref{fig:1}.
Examples of such systems include UGe$_2$ \cite{Saxena_et_al_2000} and
MnSi.\cite{Pfleiderer_et_al_1997} The equilibrium quantum phase transitions in
these systems have been studied experimentally in some detail, and our
predictions for the phase ordering should be amenable to experimental checks
using similar methods.

The nature of the mode-mode coupling effects depends on whether the system is
dirty or clean, i.e., whether or not quenched disorder is present. In the clean
case, they lead to a fluctuation-induced first-order
transition,\cite{Belitz_Kirkpatrick_Vojta_1999} so the magnetization cannot be
made arbitrarily small. In the dirty case, the quantum phase transition
generically is of second order. We will focus on the
latter,\cite{disorder_footnote} where the size of the quantum effects we
consider is not limited by a nonzero minimum value of the magnetization.
Quenched disorder also ensures that the transport coefficient $\lambda$ will
remain finite even at $T=0$. The form of the dynamical equation (\ref{eq:2.1})
will thus not be modified by quantum mechanics. We will briefly discuss clean
systems later.

Effectively, the mode-mode coupling effects in a dirty system in $2<d<4$
dimensions lead to a non-local free-energy functional that contains a
$\nabla^{d-2}$ term in addition to the usual $\nabla^2$ term, or to a Gaussian
vertex\cite{Kirkpatrick_Belitz_1996, Belitz_Kirkpatrick_Vojta_2005}
\be
\Gamma({\bm k}) = r + {\tilde c}\,\vert{\bm k}\vert^{d-2} + c\,{\bm k}^2
\label{eq:3.1}
\ee
instead of Eq.\ (\ref{eq:2.2b}), with ${\tilde c}>0$. In the ordered phase,
this nonanalyticity is cut off by the magnetization, with $m_0 \sim {\bm k}^2$.
It is also cut off by $T>0$. We will discuss the latter effect, as well as the
behavior at criticality, below; for now we assume any length scale associated
with temperature to be the largest scale in the system, which effectively sets
$T=0$, and a quench well into the ordered phase, see trajectory A in Fig.\
\ref{fig:1}.\cite{trajectory_footnote} Let us denote the length scale where
$m_0$ cuts off the nonanalyticity by $L_2(m_0) \propto m_0^{-1/2}$, and the
length scale beyond which the nonanalyticity dominates by $L^* = (c/{\tilde
c})^{1/(4-d)}$. (We will determine and discuss $L_2$ and $L^*$ in more detail
below.) Since the dissipative term and the torque term in Eq.\ (\ref{eq:2.1})
are both proportional to $\delta H/\delta{\bm\phi}$, they are equally affected.
Equation (\ref{eq:3.1}) implies that, effectively, $\delta H/\delta{\bm\phi} =
-c\nabla^2{\bm\phi}$ which would result from Eq.\ (\ref{eq:2.2a}) gets
multiplied by a function $f(\nabla,m_0)$, where $\nabla$ stands for the
appropriate inverse length scale, which is $\vert{\bm k}\vert$ for the magnon
dispersion, and $1/L$ for the phase ordering problem. For scales larger than
$L^*$, we have
\be f(\nabla,m_0) \propto \begin{cases} (L^*\,\nabla)^{d-4}
\propto \nabla^{d-4} &
                                                  \text{if}\ 1/\nabla\ll L_2 \cr
                                 (L^*/L_2)^{d-4} \propto m_0^{-(4-d)/2}
                                                    & \text{if}\ 1/\nabla\gg L_2\ .
                   \end{cases}
\label{eq:3.2}
\ee

For an illustration of these effects, let us consider the magnon dispersion
relation. The above considerations result in Eq.\ (\ref{eq:2.4}) with a
modified $D(m_0)$, viz., $D(m_0) \propto m_0^{(d-2)/2}$, for $\vert{\bm k}\vert
\ll 1/L_2$, and in $\omega({\bm k}) \propto \vert{\bm k}\vert^{d-2}$ for
$\vert{\bm k}\vert \gg 1/L_2$. The former result was first obtained in Ref.\
\onlinecite{Belitz_et_al_1998} from microscopic considerations, and we have
reproduced it here as an illustrative check on our power-counting technique.

\subsection{The quantum phase ordering problem}
\label{III.B}

For the phase ordering problem, the right-hand side of Eq.\ (\ref{eq:2.5}) gets
multiplied by $f(1/L,m_0)$,
\be
\frac{dL}{dt} \sim \left(\frac{c\lambda}{L^3} +
\frac{c\,\Omega_{\text{L}}}{L}\right)\,f(1/L,m_0).
\label{eq:3.3}
\ee
In $d=3$, the domain growth then displays four different power laws in
different time or length regimes, as follows:
\be
L(t) \propto
\begin{cases} (c\lambda/L^*)^{1/3}\,t^{1/3} & \text{if}\
                                       L<\{L_1,L_2\}\cr
                            (c\Omega_{\text{L}}/L^*)\,t & \text{if}\ L_1<L<L_2
                                                             \cr
                            (c\lambda L_2/L^*)^{1/4}\,t^{1/4} \cr \hskip 20pt
                               \propto m_0^{-1/8}\,t^{1/4} & \text{if}\ L_2<L<L_1 \cr
                            (c\Omega_{\text{L}} L_2/L^*)^{1/2}\,t^{1/2} \cr
                             \hskip 35pt \propto m_0^{1/4} t^{1/2} & \text{if}\ \{L_1,L_2\}<L
                                             \ .
             \end{cases}
\label{eq:3.4}
\ee
Compared to Eq.\ (\ref{eq:2.6}), the asymptotic time dependence of $L$ remains
unchanged, but the dependence of the pre\-factor on the equilibrium
magnetization is $m_0^{1/4}$ instead of $m_0^{1/2}$. In the initial scaling
regime, where $L<L_1$, the time dependence is $t^{1/3}$ instead of $t^{1/4}$ in
the classical case. In addition, there is an intermediate regime where $L(t)$
grows as $t$ if $L_1 < L_2$, and as $t^{1/4}$ if $L_2 < L_1$.

Equation (\ref{eq:3.4}) is the central new result of the present paper. It has
been derived entirely by power counting, that is, from Eq.\ (\ref{eq:3.3})
which associated all gradients in the problem with powers of $1/L$. Next we
establish the validity of this procedure by means of a renormalization-group
analysis that generalizes Bray's analysis of Model B.\cite{Bray_1990}

\subsection{Renormalization-group considerations}
\label{subsec:III.C}

Adapting the renormalization procedure of Ma,\cite{Ma_1976} we assign a scale
dimension $[L] = -1$ to $L$, and a scale dimension $[t] = -z$ to time. Equation
(\ref{eq:1.1}) suggests to choose the field ${\bm\phi}({\bm x},t)$ to be
dimensionless, $[{\bm\phi}({\bm x},t)]=0$. Let the Fourier transform of $C$ in
Eq.\ (\ref{eq:1.1}) be $S({\bm k},t) = \int d{\bm x}\,\exp(-i{\bm k}\cdot{\bm
x})\,C(r,t)$. The exponent $\eta$ is defined by the structure factor $S({\bm
k}) = S({\bm k},t\to\infty)$ to behave as $S({\bm k}) \propto \vert{\bm
k}\vert^{-2+\eta}$; this implies $\eta = 2-d$. Position, time, and fields are
then rescaled in the RG process according to ${\bm x}' = {\bm x}/b$, $t' =
t/b^z$, and ${\bm\phi}'({\bm x}',t') = {\bm\phi}({\bm x},t)$, respectively,
with $b$ the RG length rescaling factor. The free energy $H$, which has a naive
scale dimension equal to zero, is assigned an anomalous scale dimension $-y$,
$H'[{\bm\phi}'] = b^{-y} H[{\bm\phi}]$. Finally, one needs to keep in mind that
the functional derivative of $H$ in Eq.\ (\ref{eq:2.1}) removes a spatial
integral and therefore acts, for scaling purposes, like an inverse volume with
a scale dimension of $d$. A zero-loop renormalization of Eq.\ (\ref{eq:2.1})
then yields
\bse
\label{eqs:3.5}
\be \frac{\partial{\bm\phi}'}{\partial t'} =
\lambda'\,\nabla'^{\,2}\,\frac{\delta H'}{\delta{\bm\phi}'} +
\gamma{\,'}\,{\bm\phi}'\times\frac{\delta H'}{\delta{\bm\phi}'}.
\label{eq:3.5a}
\ee
with renormalized quantities
\bea
\lambda' &=& \lambda\,b^{z-d-2+y},
\nonumber\\
\gamma{\,'} &=& \gamma\,b^{z-d+y}.
\label{eq:3.5b}
\eea
\ese

For Model B ($\gamma=0$), the assumption that the transport coefficient
$\lambda$ is not singularly renormalized at the fixed point we are looking for
(which assumes that a hydrodynamic description remains valid in the ordered
phase), leads to the relation $z = d+2-y$,\cite{Bray_1990} which expresses the
dynamical exponent $z$ in terms of the energy exponent $y$. For Model J this
fixed point is not stable: $\gamma$ is relevant with respect to it. Assuming
that $\gamma$ is not singularly renormalized (which assumes that the spin waves
in the ordered phase are characterized by $\omega({\bm k}) \propto {\bm k}^2$),
leads to
\be
z = d-y.
\label{eq:3.6}
\ee
The remaining question is the value of the anomalous energy dimension $y$. If
defects in the order parameter texture determine the scaling properties of the
energy, then $y=d-2$ for a vector order parameter.\cite{Bray_1990} This yields
$z=2$, in agreement with the long-time behavior of $L(t)$. For Model B, the
same value of $y$ yields $z=4$.\cite{Bray_1990} With increasing length scale,
one thus expects a crossover from $z=4$ to $z=2$, as is reflected in Eq.\
(\ref{eq:3.4}). If $L < L_2$, then one effectively has $\nabla^{d-2}$ in the
free energy instead of $\nabla^2$, so one expects $y = d - (d-2) = 2$. This
leads to $z=d$ for Model B, and $z=d-2$ for Model J, as reflected in the first
two lines in Eq.\ (\ref{eq:3.4}). The above considerations show that the naive
power-counting considerations that lead to Eq.\ (\ref{eq:2.5}) or
(\ref{eq:3.3}), which replace all gradients in the dynamical equation by $1/L$,
are indeed correct, subject to the above assumptions. Note that for an Ising
order parameter, $y=d-1$,\cite{Bray_1990} and hence $z=3$ for Model B, so the
naive power counting breaks down.\cite{Ising_footnote}

\section{Discussion and Conclusion}
\label{sec:IV}

In the remainder of the paper we provide a semi-quantitative discussion of Eq.\
(\ref{eq:3.4}) by identifying the various length scales that enter the quantum
phase ordering problem. $L_1$ has been identified in the context of Eq.\
(\ref{eq:2.6}). $L_2$ can be identified from the explicit treatment of the
ferromagnetic phase in Ref.\ \onlinecite{Sessions_Belitz_2003}. We find $L_2 =
\sqrt{D/\Delta}$, where $D$ is the charge diffusion constant and $\Delta$ is
the Stoner gap or exchange splitting. A related scale is $L_T = \sqrt{D/T}$,
which denotes the length scale where a nonzero temperature cuts off the
nonanalyticity. Finally, $L^*$ was introduced in connection with Eq.\
(\ref{eq:3.2}). Reference\ \onlinecite{Belitz_et_al_2001a} yields explicit
expressions for the coefficients $c$ and ${\tilde c}$, which give $L^* =
\pi\ell/72$, with $\ell$ the elastic electronic mean-free path due to the
quenched disorder. Thus,
\bea
L_1 &=& \sqrt{\lambda/\Omega_{\text{L}}}\quad,\quad L_2 = \sqrt{D/\Delta}\ ,
\nonumber\\
L_T &=& \sqrt{D/T}\hskip 13pt ,\quad L^* = \pi\ell/72\ .
\label{eq:4.1}
\eea

We now estimate the values of these scales. $\Omega_{\text{L}}$ is on the order
of $em_0/2m_{\text{e}}c$, with $e$ and $m_{\text{e}}$ the electron charge and
mass, respectively, and $c$ the speed of light (not to be confused with the
coefficient of the square gradient term in the Hamiltonian that is denoted by
$c$ everywhere else in this paper). The effects we are considering are largest
if the magnetization is small; either because the system is a weak magnet, or
because the quench is to just within the ordered phase (but outside the
critical region; we consider a critical quench below). For a magnetization $m_0
= 10 - 100\,{\text{G}}$ we have
$\hbar\Omega_{\text{L}}/k_{\text{B}} \approx 0.001 - 0.01\,\text{K}$. For the
Stoner gap one expects $\hbar\Delta/k_{\text{B}} \agt T_{\text{c}}$, with
$T_{\text{c}}$ the critical temperature that corresponds to the parameter
values after the quench. For low-$T_{\text c}$ magnets like MnSi or UGe$_2$,
this means $\hbar\Delta/k_{\text{B}} \approx 10^2\,{\text{K}}$. With
free-electron parameters, and a Fermi wave number $k_{\text{F}} \approx
1\,\AA^{-1}$, the mean-free path is related to the resistivity $\rho$ by $\ell
\approx 10^3 (\mu\Omega\text{cm}/\rho)\,\AA$, and in the ordered phase one
expects $\lambda \approx D \approx \hbar k_{\text{F}}\ell/3m_{\text{e}}$. For
$\rho = 10\,\mu\Omega\text{cm}$ and $T=1\,{\text{K}}$, a rough estimate for the
hierarchy of length scales thus is
\bea
L^* &\approx& 5\,\AA\hskip 19pt ,\quad L_2 \approx 10^2\,\AA\ ,
\nonumber\\
L_T &\approx& 10^3\,\AA\quad,\quad L_1 \approx 10^5\,\AA\ .
\label{eq:4.2}
\eea
For $L^* < L < L_2$ one has the initial $t^{1/3}$ behavior in Eq.\
(\ref{eq:3.4}). For $L_2 < L < L_1$, the domain size will grow as $L(t) \propto
m_0^{-1/8}\,t^{1/4}$, and for $L > L_1$ the behavior crosses over to $L(t)
\propto m_0^{1/4}\,t^{1/2}$. With respect to the latter, one should keep in
mind that domains larger than a few tens of microns are hard to achieve in zero
magnetic field, except close to the critical
point.\cite{Landau_Lifshitz_VIII_1984} These predictions should be observable
by time-resolved neutron scattering. In particular, the magnetization
dependence of the prefactor of the asymptotic $t^{1/2}$ law can be checked by
quenching along trajectory A in Fig.\ \ref{fig:1} to different final pressure
values. By recalling that the parameter $c$ in Eq.\ (\ref{eq:2.2a}) represents
the square of a microscopic length scale that is on the order of an $\AA$ we
can estimate the time required for a domain to grow to sizes corresponding to
the various length scales given in Eq.\ (\ref{eq:4.2}):
\bea
t^* &\approx& 10^{-15}\,{\rm s}\hskip 6pt ,\quad t_2 \approx 10^{-11}\,{\rm s}\
,
\nonumber\\
t_T &\approx& 10^{-7}\,{\rm s}\quad,\quad t_1 \approx 1\,{\rm s}\ .
\label{eq:4.3}
\eea
Notice that the microscopic time scale for the problem is given by the Fermi
wave length divided by the Fermi velocity, which is about $10^{-15}\,{\rm s}$
with free-electron parameters. This is consistent with the value of $t^*$.

Now consider a quench into the critical region, which is divided into regimes
denoted by I, II, and III in Fig.\ (\ref{fig:1}). The classical critical fixed
point controls Region I, where phase ordering has been discussed by Das and
Rao.\cite{Das_Rao_2000} Regions II and III are controlled by the quantum
critical fixed point, and the quantum critical behavior is known
exactly.\cite{Belitz_et_al_2001b} Consider a quench to the quantum critical
point, trajectory B in the figure.\cite{trajectory_footnote} The quantum
ferromagnetic critical behavior is characterized by logarithmic corrections to
scaling, which can be expressed in terms of scale dependent critical exponents.
The dynamical critical exponent in $d=3$ is \cite{Belitz_et_al_2001b,
Belitz_Kirkpatrick_Vojta_2005}
\be z = 3 + \text{const.}\times(\ln\ln b)^2/\ln b,
\label{eq:4.4}
\ee
to leading logarithmic accuracy. At the quantum critical point, and in the
context of domain growth, the renormalization-group scale factor $b$ represents
$1/L(t)$. This leads to a growth law, with the time $t$ measured in arbitrary
units,
\be L(t\to\infty) \propto t^{1/3}\,e^{\text{const.}\times(\ln\ln t)^2}.
\label{eq:4.5}
\ee

If the quench ends at a low temperature in the critical region, but not at the
quantum critical point, $L(t)$ will grow according to Eq.\ (\ref{eq:4.5}) until
it becomes comparable to the correlation length $\xi$. In region IIb, at longer
times it will saturate at a value comparable to $\xi$, while in region IIa
there will be a crossover to the asymptotic behavior as described by Eq.\
(\ref{eq:3.4}). A more complete description of critical quenches will be given
elsewhere.\cite{us_tbp}

In clean systems, analogous mode-mode coupling effects lead to a weaker
nonanalytic term than in Eq.\ (\ref{eq:3.1}); in $d=3$ it is ${\bm
k}^2\ln\vert{\bm k}\vert$. However, the term is {\em negative}, which leads to
a first-order transition.\cite{Belitz_Kirkpatrick_Vojta_1999} The order of the
transition is of no consequence for the phase ordering kinetics, but the
requirement of a positive transverse magnetic susceptibility in the ordered
phase prevents the magnetization from ever being small enough for the
nonanalytic term to dominate over the analytic one. For the magnon dispersion
relation, one finds $D(m_0) = \gamma\,c\,m_0[1 -
\text{const.}\times\ln(1/m_0)]$, and the equation of state will ensure that
$D(m_0) > 0$.\cite{magnons_footnote} Similarly, for the phase ordering problem
one has $L(t) \propto [1 - \text{const.}\times\ln t]\,t^{1/3}$ in a transient
regime, and $L(t\to\infty) \propto m_0^{1/2}\,[1 -
\text{const.}\times\ln(1/m_0)]\,t^{1/2}$ asymptotically (const.$\,>0$). The
former result follows since, in the clean limit at $T=0$, $\lambda({\bm k}\to
0) \propto 1/\vert{\bm k}\vert$.

We conclude by summarizing the original results obtained in this paper. First,
we generalized the phenomenology that was developed to describe phase ordering
following a quench across classical phase transitions to the quantum phase
transition case. Second, for the continuous quantum phase transition expected
in disordered Heisenberg quantum ferromagnets we obtained the growth laws in
various time windows for quenches both deep into the ordered phase and to a
point at or very near quantum criticality. Third, we gave the domain growth
laws for clean itinerant Heisenberg quantum magnets, where the quantum phase
transition is expected to be discontinuous. In the latter case the quantum
effects are subleading due to the lower bound on the magnetization imposed by
the first-order nature of the transition. All of these results are amenable to
experimental verification.

\acknowledgments
We thank Dave Cohen for a useful discussion. This work was
supported by the NSF under grant Nos. DMR-05-29966 and DMR-05-30314.


\begin{thebibliography}{29}
\expandafter\ifx\csname natexlab\endcsname\relax\def\natexlab#1{#1}\fi
\expandafter\ifx\csname bibnamefont\endcsname\relax
  \def\bibnamefont#1{#1}\fi
\expandafter\ifx\csname bibfnamefont\endcsname\relax
  \def\bibfnamefont#1{#1}\fi
\expandafter\ifx\csname citenamefont\endcsname\relax
  \def\citenamefont#1{#1}\fi
\expandafter\ifx\csname url\endcsname\relax
  \def\url#1{\texttt{#1}}\fi
\expandafter\ifx\csname urlprefix\endcsname\relax\def\urlprefix{URL }\fi
\providecommand{\bibinfo}[2]{#2}
\providecommand{\eprint}[2][]{\url{#2}}

\bibitem[{\citenamefont{Hohenberg and
  Halperin}(1977)}]{Hohenberg_Halperin_1977}
\bibinfo{author}{\bibfnamefont{P.~C.} \bibnamefont{Hohenberg}}
  \bibnamefont{and} \bibinfo{author}{\bibfnamefont{B.~I.}
  \bibnamefont{Halperin}}, \bibinfo{journal}{Rev. Mod. Phys.}
  \textbf{\bibinfo{volume}{49}}, \bibinfo{pages}{435} (\bibinfo{year}{1977}).

\bibitem[{\citenamefont{Zurek}(1985)}]{Zurek_1996}
\bibinfo{author}{\bibfnamefont{W.~H.} \bibnamefont{Zurek}},
  \bibinfo{journal}{Phys. Rep.} \textbf{\bibinfo{volume}{276}},
  \bibinfo{pages}{177} (\bibinfo{year}{1985}).

\bibitem[{dom()}]{domains_footnote}
\bibinfo{note}{We will consider Heisenberg magnets, which do not have domains
  in the same sense as Ising magnets. By ``domain size'' we mean the linear
  size of a region in space over which the local magnetization points on
  average in a given direction.}

\bibitem[{\citenamefont{Bray}(1994)}]{Bray_1994}
\bibinfo{author}{\bibfnamefont{A.~J.} \bibnamefont{Bray}},
  \bibinfo{journal}{Adv. Phys.} \textbf{\bibinfo{volume}{43}},
  \bibinfo{pages}{357} (\bibinfo{year}{1994}).

\bibitem[{\citenamefont{Sachdev}(1999)}]{Sachdev_1999}
\bibinfo{author}{\bibfnamefont{S.}~\bibnamefont{Sachdev}},
  \emph{\bibinfo{title}{Quantum Phase Transitions}}
  (\bibinfo{publisher}{Cambridge University Press, Cambridge},
  \bibinfo{year}{1999}).

\bibitem[{\citenamefont{Belitz et~al.}(2005)\citenamefont{Belitz, Kirkpatrick,
  and Vojta}}]{Belitz_Kirkpatrick_Vojta_2005}
\bibinfo{author}{\bibfnamefont{D.}~\bibnamefont{Belitz}},
  \bibinfo{author}{\bibfnamefont{T.~R.} \bibnamefont{Kirkpatrick}},
  \bibnamefont{and} \bibinfo{author}{\bibfnamefont{T.}~\bibnamefont{Vojta}},
  \bibinfo{journal}{Rev. Mod. Phys.} \textbf{\bibinfo{volume}{77}},
  \bibinfo{pages}{579} (\bibinfo{year}{2005}).

\bibitem[{\citenamefont{Ma and Mazenko}(1975)}]{Ma_Mazenko_1975}
\bibinfo{author}{\bibfnamefont{S.-K.} \bibnamefont{Ma}} \bibnamefont{and}
  \bibinfo{author}{\bibfnamefont{G.~F.} \bibnamefont{Mazenko}},
  \bibinfo{journal}{Phys. Rev. B} \textbf{\bibinfo{volume}{11}},
  \bibinfo{pages}{4077} (\bibinfo{year}{1975}).

\bibitem[{\citenamefont{Ma}(1976)}]{Ma_1976}
\bibinfo{author}{\bibfnamefont{S.-K.} \bibnamefont{Ma}},
  \emph{\bibinfo{title}{Modern Theory of Critical Phenomena}}
  (\bibinfo{publisher}{Benjamin, Reading, MA}, \bibinfo{year}{1976}).

\bibitem[{Isi()}]{Ising_footnote}
\bibinfo{note}{This assumption is correct for the Heisenberg model, but not for
  an Ising model. In the latter case, a microscopic length, namely, the domain
  wall thickness $\xi$, enters in addition to the length scale $L$.
  Effectively, one then has $\nabla^3 \sim 1/L^2\xi$,\cite{notation_footnote}
  and the dynamical exponent is $z=3$. In a renormalization-group language,
  $\xi$ is a dangerously irrelevant operator. See Ref.\ \onlinecite{Bray_1994}
  for a detailed discussion.}

\bibitem[{not()}]{notation_footnote}
\bibinfo{note}{We use the symbols $\sim$ and $\propto$ to mean ``equal for
  scaling or power-counting purposes'', and ``proportional to'', respectively.}

\bibitem[{gra()}]{gradients_footnote}
\bibinfo{note}{In equilibrium, the terms in $\delta H/\delta{\bm\phi}$ without
  gradients vanish on average. Upon approaching equilibrium, they are therefore
  small and do not change the behavior of $L(t\to\infty)$. They do, however,
  produce a power-law pre\-factor to the exponential decay of the pair
  correlation function that is important to ensure the proper scaling behavior
  of the latter, Eq.\ (\ref{eq:1.1}). This can be seen explicitly in a
  large-$n$ solution of the problem, see Ref.\ \onlinecite{Bray_1994}.}

\bibitem[{\citenamefont{Moriya}(1985)}]{Moriya_1985}
\bibinfo{author}{\bibfnamefont{T.}~\bibnamefont{Moriya}},
  \emph{\bibinfo{title}{Spin Fluctuations in Itinerant Electron Magnetism}}
  (\bibinfo{publisher}{Springer, Berlin}, \bibinfo{year}{1985}).

\bibitem[{\citenamefont{Das and Rao}(2000)}]{Das_Rao_2000}
\bibinfo{author}{\bibfnamefont{J.}~\bibnamefont{Das}} \bibnamefont{and}
  \bibinfo{author}{\bibfnamefont{M.}~\bibnamefont{Rao}},
  \bibinfo{journal}{Phys. Rev. E} \textbf{\bibinfo{volume}{62}},
  \bibinfo{pages}{1601} (\bibinfo{year}{2000}).

\bibitem[{\citenamefont{Kirkpatrick and
  Belitz}(1996)}]{Kirkpatrick_Belitz_1996}
\bibinfo{author}{\bibfnamefont{T.~R.} \bibnamefont{Kirkpatrick}}
  \bibnamefont{and} \bibinfo{author}{\bibfnamefont{D.}~\bibnamefont{Belitz}},
  \bibinfo{journal}{Phys. Rev. B} \textbf{\bibinfo{volume}{53}},
  \bibinfo{pages}{14364} (\bibinfo{year}{1996}).

\bibitem[{\citenamefont{Belitz et~al.}(2001{\natexlab{a}})\citenamefont{Belitz,
  Kirkpatrick, Mercaldo, and Sessions}}]{Belitz_et_al_2001a}
\bibinfo{author}{\bibfnamefont{D.}~\bibnamefont{Belitz}},
  \bibinfo{author}{\bibfnamefont{T.~R.} \bibnamefont{Kirkpatrick}},
  \bibinfo{author}{\bibfnamefont{M.~T.} \bibnamefont{Mercaldo}},
  \bibnamefont{and} \bibinfo{author}{\bibfnamefont{S.}~\bibnamefont{Sessions}},
  \bibinfo{journal}{Phys. Rev. B} \textbf{\bibinfo{volume}{63}},
  \bibinfo{pages}{174427} (\bibinfo{year}{2001}{\natexlab{a}}).

\bibitem[{\citenamefont{Hertz}(1976)}]{Hertz_1976}
\bibinfo{author}{\bibfnamefont{J.}~\bibnamefont{Hertz}},
  \bibinfo{journal}{Phys. Rev. B} \textbf{\bibinfo{volume}{14}},
  \bibinfo{pages}{1165} (\bibinfo{year}{1976}).

\bibitem[{\citenamefont{Belitz et~al.}(2001{\natexlab{b}})\citenamefont{Belitz,
  Kirkpatrick, Mercaldo, and Sessions}}]{Belitz_et_al_2001b}
\bibinfo{author}{\bibfnamefont{D.}~\bibnamefont{Belitz}},
  \bibinfo{author}{\bibfnamefont{T.~R.} \bibnamefont{Kirkpatrick}},
  \bibinfo{author}{\bibfnamefont{M.~T.} \bibnamefont{Mercaldo}},
  \bibnamefont{and} \bibinfo{author}{\bibfnamefont{S.}~\bibnamefont{Sessions}},
  \bibinfo{journal}{Phys. Rev. B} \textbf{\bibinfo{volume}{63}},
  \bibinfo{pages}{174428} (\bibinfo{year}{2001}{\natexlab{b}}).

\bibitem[{\citenamefont{Belitz et~al.}(1998)\citenamefont{Belitz, Kirkpatrick,
  Millis, and Vojta}}]{Belitz_et_al_1998}
\bibinfo{author}{\bibfnamefont{D.}~\bibnamefont{Belitz}},
  \bibinfo{author}{\bibfnamefont{T.~R.} \bibnamefont{Kirkpatrick}},
  \bibinfo{author}{\bibfnamefont{A.~J.} \bibnamefont{Millis}},
  \bibnamefont{and} \bibinfo{author}{\bibfnamefont{T.}~\bibnamefont{Vojta}},
  \bibinfo{journal}{Phys. Rev. B} \textbf{\bibinfo{volume}{58}},
  \bibinfo{pages}{14155} (\bibinfo{year}{1998}).

\bibitem[{\citenamefont{Saxena et~al.}(2000)\citenamefont{Saxena, Agarwal,
  Ahilan, Grosche, Haselwimmer, Steiner, Pugh, Walker, Julian, Monthoux
  et~al.}}]{Saxena_et_al_2000}
\bibinfo{author}{\bibfnamefont{S.~S.} \bibnamefont{Saxena}},
  \bibinfo{author}{\bibfnamefont{P.}~\bibnamefont{Agarwal}},
  \bibinfo{author}{\bibfnamefont{K.}~\bibnamefont{Ahilan}},
  \bibinfo{author}{\bibfnamefont{F.~M.} \bibnamefont{Grosche}},
  \bibinfo{author}{\bibfnamefont{R.~K.~W.} \bibnamefont{Haselwimmer}},
  \bibinfo{author}{\bibfnamefont{M.~J.} \bibnamefont{Steiner}},
  \bibinfo{author}{\bibfnamefont{E.}~\bibnamefont{Pugh}},
  \bibinfo{author}{\bibfnamefont{I.~R.} \bibnamefont{Walker}},
  \bibinfo{author}{\bibfnamefont{S.~R.} \bibnamefont{Julian}},
  \bibinfo{author}{\bibfnamefont{P.}~\bibnamefont{Monthoux}},
  \bibnamefont{et~al.}, \bibinfo{journal}{Nature}
  \textbf{\bibinfo{volume}{406}}, \bibinfo{pages}{587} (\bibinfo{year}{2000}).

\bibitem[{\citenamefont{Pfleiderer et~al.}(1997)\citenamefont{Pfleiderer,
  McMullan, Julian, and Lonzarich}}]{Pfleiderer_et_al_1997}
\bibinfo{author}{\bibfnamefont{C.}~\bibnamefont{Pfleiderer}},
  \bibinfo{author}{\bibfnamefont{G.~J.} \bibnamefont{McMullan}},
  \bibinfo{author}{\bibfnamefont{S.~R.} \bibnamefont{Julian}},
  \bibnamefont{and} \bibinfo{author}{\bibfnamefont{G.~G.}
  \bibnamefont{Lonzarich}}, \bibinfo{journal}{Phys. Rev. B}
  \textbf{\bibinfo{volume}{55}}, \bibinfo{pages}{8330} (\bibinfo{year}{1997}),
  \bibinfo{note}{$ $MnSi is actually a weak helimagnet}.

\bibitem[{\citenamefont{Belitz et~al.}(1999)\citenamefont{Belitz, Kirkpatrick,
  and Vojta}}]{Belitz_Kirkpatrick_Vojta_1999}
\bibinfo{author}{\bibfnamefont{D.}~\bibnamefont{Belitz}},
  \bibinfo{author}{\bibfnamefont{T.~R.} \bibnamefont{Kirkpatrick}},
  \bibnamefont{and} \bibinfo{author}{\bibfnamefont{T.}~\bibnamefont{Vojta}},
  \bibinfo{journal}{Phys. Rev. Lett.} \textbf{\bibinfo{volume}{82}},
  \bibinfo{pages}{4707} (\bibinfo{year}{1999}).

\bibitem[{dis()}]{disorder_footnote}
\bibinfo{note}{We will consider only average effects of the quenched disorder.
  For a more complete numerical study of domain growth in an Ising system with
  quenched disorder, see, Ref. \onlinecite{Paul_Puri_Rieger_2005}.}

\bibitem[{tra()}]{trajectory_footnote}
\bibinfo{note}{Figure\ \ref{fig:1} shows a pure pressure quench and a pure
  temperature quench, respectively. The growth law depends only on the final
  state, not on the quench trajectory.}

\bibitem[{\citenamefont{Bray}(1990)}]{Bray_1990}
\bibinfo{author}{\bibfnamefont{A.~J.} \bibnamefont{Bray}},
  \bibinfo{journal}{Phys. Rev. B} \textbf{\bibinfo{volume}{41}},
  \bibinfo{pages}{6724} (\bibinfo{year}{1990}).

\bibitem[{\citenamefont{Sessions and Belitz}(2003)}]{Sessions_Belitz_2003}
\bibinfo{author}{\bibfnamefont{S.}~\bibnamefont{Sessions}} \bibnamefont{and}
  \bibinfo{author}{\bibfnamefont{D.}~\bibnamefont{Belitz}},
  \bibinfo{journal}{Phys. Rev. B} \textbf{\bibinfo{volume}{68}},
  \bibinfo{pages}{054411} (\bibinfo{year}{2003}).

\bibitem[{\citenamefont{Landau and Lifshitz}(1984)}]{Landau_Lifshitz_VIII_1984}
\bibinfo{author}{\bibfnamefont{L.~D.} \bibnamefont{Landau}} \bibnamefont{and}
  \bibinfo{author}{\bibfnamefont{E.~M.} \bibnamefont{Lifshitz}},
  \emph{\bibinfo{title}{Electrodynamics of Continuous Media}}
  (\bibinfo{publisher}{Pergamon, Oxford}, \bibinfo{year}{1984}).

\bibitem[{us_()}]{us_tbp}
\bibinfo{note}{R. Saha, T.R. Kirkpatrick, and D. Belitz, to be published.}

\bibitem[{mag()}]{magnons_footnote}
\bibinfo{note}{The functional form of the effect for the clean case as
  discussed in Ref.\ \onlinecite{Belitz_et_al_1998} was correct, but the sign
  was not, and the fact that the nonanalytic term will necessarily be smaller
  than the regular one was not mentioned.}

\bibitem[{\citenamefont{Paul et~al.}(2005)\citenamefont{Paul, Puri, and
  Rieger}}]{Paul_Puri_Rieger_2005}
\bibinfo{author}{\bibfnamefont{R.}~\bibnamefont{Paul}},
  \bibinfo{author}{\bibfnamefont{S.}~\bibnamefont{Puri}}, \bibnamefont{and}
  \bibinfo{author}{\bibfnamefont{H.}~\bibnamefont{Rieger}},
  \bibinfo{journal}{Phys. Rev. E} \textbf{\bibinfo{volume}{71}},
  \bibinfo{pages}{061109} (\bibinfo{year}{2005}).

\end{thebibliography}

\end{document}